# A RESOLVED MAP OF THE INFRARED EXCESS IN A LYMAN BREAK GALAXY AT $Z = 3$

M. P. Koprowski[1] K. E. K. Coppin[1], J. E. Geach[1], N. K. Hine[1], M. Bremer[2], S. Chapman[3], L. J. M. Davies[4], T. Hayashino[5], K. K. Knudsen[6], M. Kubo[7], B. D. Lehmer[8], Y. Matsuda[7,9], D. J. B. Smith[1], P. P. van der Werf[10], G. Violino[1], T. Yamada[11]

[1]Centre for Astrophysics Research, University of Hertfordshire, College Lane, Hatfield AL10 9AB, UK
[2]H.H. Wills Physics Laboratory, University of Bristol, Tyndall Avenue, Bristol BS8 1TL, UK
[3]Department of Physics and Atmospheric Science, Dalhousie University, Halifax, NS B3H 4R2, Canada
[4]ICRAR, The University of Western Australia, 35 Stirling Highway, Crawley, WA 6009, Australia
[5]Research Center for Neutrino Science, Graduate School of Science, Tohoku University, Sendai, Miyagi 980-8578, Japan
[6]Department of Earth and Space Sciences, Chalmers University of Technology, Onsala Space Observatory, SE-43992 Onsala, Sweden
[7]National Astronomical Observatory of Japan, Osawa 2-21-1, Mitaka, Tokyo 181-8588, Japan
[8]Department of Physics, University of Arkansas, 226 Physics Building, 835 West Dickson Street, Fayetteville, AR 72701, USA
[9]Graduate University for Advanced Studies, Osawa 2-21-1, Mitaka, Tokyo 181-8588, Japan
[10]Leiden Observatory, Leiden University, P.O. Box 9513, NL-2300 RA Leiden, The Netherlands
[11]Institute of Space and Astronautical Science, Japan Aerospace Exploration Agency, 3-1-1 Yoshinodai, 252-5210 Sagamihara, Kanagawa 252-5210, Japan

*Draft version August 12, 2016*

## ABSTRACT

We have observed the dust continuum of ten $z = 3.1$ Lyman Break Galaxies with the Atacama Large Millimeter/Submillimeter Array at ∼450 mas resolution in Band 7. We detect and resolve the 870$\mu$m emission in one of the targets with a flux density of $S_{870} = 192 \pm 57 \mu$Jy, and measure a stacked $3\sigma$ signal of $S_{870} = 67 \pm 23 \mu$Jy for the remaining nine. The total infrared luminosities are $L_{8-1000} = (8.4 \pm 2.3) \times 10^{10}$ L$_\odot$ for the detection and $L_{8-1000} = (2.9 \pm 0.9) \times 10^{10}$ L$_\odot$ for the stack. With *Hubble Space Telescope* Advanced Camera for Surveys *I*-band imaging we map the rest-frame UV emission on the same scale as the dust, effectively resolving the 'infrared excess' (IRX = $L_{\rm FIR}/L_{\rm UV}$) in a normal galaxy at $z = 3$. Integrated over the galaxy we measure IRX = $0.56 \pm 0.15$, and the galaxy-averaged UV slope is $\beta = -1.25 \pm 0.03$. This puts the galaxy a factor of ∼10 below the IRX-$\beta$ relation for local starburst nuclei of Meurer et al. (1999). However, IRX varies by more than a factor of 3 across the galaxy, and we conclude that the complex relative morphology of the dust relative to UV emission is largely responsible for the scatter in the IRX-$\beta$ relation at high-$z$. A naive application of a Meurer-like dust correction based on the UV slope would dramatically over-estimate the total star formation rate, and our results support growing evidence that when integrated over the galaxy, the typical conditions in high-$z$ star-forming galaxies are not analogous to those in the local starburst nuclei used to establish the Meurer relation.

## 1. INTRODUCTION

It is now established that the global star formation rate (SFR) density ($\rho_{\rm SFR}$) steadily declines beyond $z \approx 3$ following $\rho_{\rm SFR} \propto (1 + z)^{-6}$ (Ellis et al. 2013; Oesch et al. 2014, 2015; Bouwens et al. 2015; McLeod et al. 2016). This 'ramp up' epoch of star formation at $z \gtrsim 2$ is a new frontier of observational cosmology. However, meaningful samples of galaxies at $z > 3$ have only been possible by selecting Lyman break drop outs in ultradeep optical/near-infrared imaging, resulting in rest-frame UV selected samples. Estimates of the total SFRs of these galaxies are made by correcting the UV luminosities for internal dust extinction based on the slope, $\beta$, of the UV continuum (where $f_\lambda \propto \lambda^\beta$). Usually this is couched in terms of the 'Infrared Excess' (IRX), IRX = $L_{\rm FIR}/L_{\rm UV}$ (Meurer, Heckman, & Calzetti 1999, hereafter MHC99), with IRX related to $\beta$ in a manner that assumes that the UV/optical photons are absorbed by interstellar dust, increasing $\beta$, are re-emitted in the far-infrared[1].

The IRX-$\beta$ relation is useful because, at high-$z$, one typically only has a direct measurement of $L_{\rm UV}$ (usually defined as $\nu L_{1600}$) and $\beta$. Thus, given an uncorrected $L_{\rm UV}$ and $\beta$ it is possible to predict $L_{\rm FIR}$ and therefore the total SFR. A simple question is whether or not the same IRX-$\beta$ correction derived for local systems can be applied at high-$z$. There are several reasons for worry: (1) star formation might be proceeding in a different manner in the gas-rich discs of early galaxies compared to (comparatively) quiescent local discs and starburst nuclei (Tacconi et al. 2010; Swinbank et al. 2015), (2) the evolution of dust production at early times has not yet been established; it is certainly expected that systematic differences in the metallicity of the interstellar medium (ISM) of high-$z$ galaxies compared to local systems could result in a different dust reddening law, (3) the IR and UV emitting regions might not be spatially coincident - there could be both heavily obscured and unobscured lines of sight in the same source (Douglas et al. 2009), which will affect galaxy-averaged measurements of IRX.

It is now possible to directly detect the far-infrared dust emission of 'normal' star-forming galaxies at high-$z$. Recent observations are revealing an interesting, but confusing, picture: Capak et al. (2015) measured the IRX for $z \sim 5$ LBGs detected at far-infrared and found them to have significantly lower values than similar sources in the local Universe. This indicates that the MHC99 relation may not necessarily hold at high-$z$. However, Watson et al. (2015), report the detection of dust emission from a (lensed) galaxy with a rather blue UV continuum at $z = 7.5$, finding an IRX to be consistent with the local relation for the measured $\beta$.

---
[1] Note that in the original definition $L_{\rm FIR}$ is the far-infrared luminosity defined by Helou et al. (1988) from *IRAS* band passes, not the integrated 8-1000$\mu$m luminosity, as is often used in the literature, which is ∼50% higher for a typical dust spectrum.

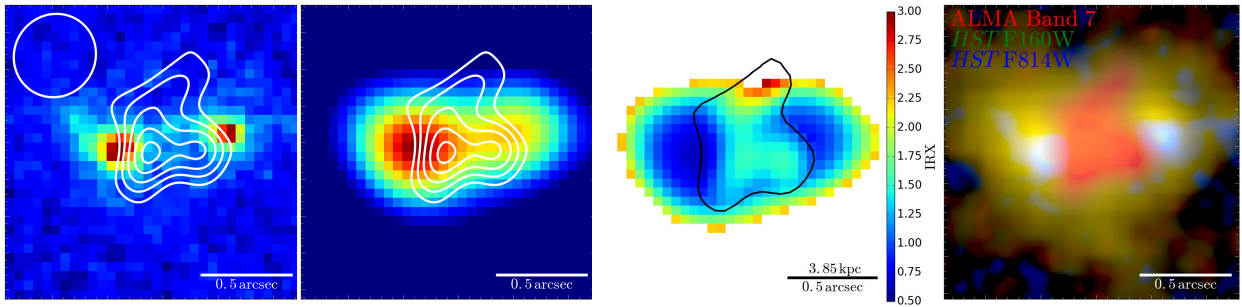

**Figure 1.** **Panel 1:** The ALMA 870μm contours overlaid on the *HST* F814W image. The contours are 2, 2.5, 3, 3.5 and 4σ. The ALMA synthesized beam of $0.46'' \times 0.44''$ (PA=-33.32 degrees) is shown in the top left corner. **Panel 2:** The *HST* image was convolved with the ALMA synthesized beam of our observations in order to construct a resolved IRX map **Panel 3:** The IRX map with a pixel size of $0.05''$. The ALMA flux at each pixel was translated to $L_{\rm FIR}$ using the best-fit SED from the top panel in Figure 3. Similarly, the *HST* flux at each pixel was translated to $L_{\rm UV}$ using the same SED. The IRX peaks at $\simeq 1.5$ and decreases to $\lesssim 0.5$ across the LBG, with values outside the contour being 2σ upper limits. **Panel 4:** The RGB plot for SSA22a-C16 with red, green and blue channels representing ALMA Band 7, *HST*/WFC3 F160W and *HST*/ACS F814W bands respectively.

In this Letter we present new 870μm observations of Lyman Break Galaxies at $z = 3$ using the Atacama Large Millimeter/Submillimeter Array (ALMA). Combined with *HST* ACS imaging we resolve the IRX in a single target and thereby investigate the relative spatial extent of the stellar emission and dust absorption and the implication it has on the integrated value of the IRX. We assume a flat cosmology with $\Omega_m = 0.3, \Omega_\Lambda = 0.7$ and $H_0 = 70$ km s$^{-1}$Mpc$^{-1}$.

## 2. DATA AND OBSERVATIONS

### 2.1. *Targets*

Our sample is taken from the Steidel et al. (2003) LBG redshift survey of the SSA22 field using the Palomar 5.08-m telescope. The optical magnitudes of the targets span $24 \leq R_{AB} \leq 26$ mag, and spectroscopic redshifts have been obtained using the Low Resolution Imaging Spectrometer (LRIS) on Keck (Oke et al. 1995; see Table 1).

### 2.2. *Atacama Large Millimeter/Submillimeter Array*

Coppin et al. (2015) have measured average 850μm (stacked) flux densities for LBGs at $z = 3$–5, with canonically selected LBGs typically $S_{850} \approx 250$ μJy at $z \approx 3$ (although with a clear mass dependence on the stacked flux density). This guided the sensitivity requirements of the follow-up ALMA Band 7 continuum observations we present here: the target 1σ noise for a $1''$ beam was $\sigma = 50$ μJy beam$^{-1}$.

The ALMA Band 7 observations were taken between 2014 June 30 and 2015 April 29 as a part of Cycle 2 Project #2013.1.00362.S. The observations were split into four scheduling blocks (SBs) with each target being observed for 1073 seconds. The antenna configuration delivered a resolution of approximately 500 mas and the programme was deemed complete when the depth reached 50 μJy beam$^{-1}$. Unfortunately, this means that we are less sensitive to submillimeter emission in the LBGs if the dust is extended on scales larger than the delivered ~500 mas beam. As we will show in this Section, the dust emission is resolved on this scale.

Neptune or quasar J2148+0657 were used as flux calibrators (with a 5-10% uncertainty on the absolute flux calibration), and the quasar J2148+0657 was used as a phase calibrator for all SBs. The data were reduced and imaged using the Common Astronomy Software Application (CASA) version 4.4.0[2]. Calibration involved first applying the system temperature and phase corrections and initial flagging of the autocorrelation data (shadowed and noisy antennas, channel edges, etc.). The bandpass calibration was then performed, with the phase-only gain solutions applied on-the-fly (with only central spectral channels used). This step ensures that the bandpass calibrator's temporal phase variations are correct. The bandpass solution was then applied during the gain calibrations. The resulting solutions were then visually examined, with any problematic regions flagged, and applied to the science fields.

[2] http://casa.nrao.edu

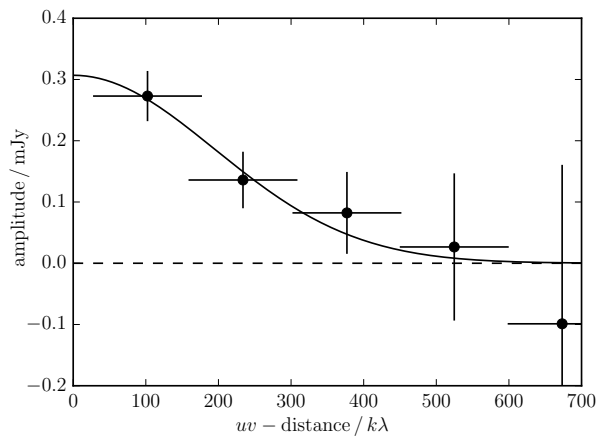

**Figure 2.** Average amplitude versus *uv*-distance for SSA22a-C16, evaluated in bins of 150 kλ. An unresolved source has a constant amplitude for all baselines, but the data are better fit by a Gaussian (FWHM $0.94''$), indicating that we have resolved the dust emission in this source.

To image the visibilities we used the CASA *clean* task with natural weighting to maximize the signal-to-noise. Only one target is detected (SSA22a-C16 at $z = 3.065$) with a peak significance of 4σ. We show the ALMA detection in Figure 1, contoured (2–4σ) on the *HST* F814W image. The astrometry of the *HST* image was registered to the ground-based Subaru imaging with an r.m.s. error of $0.2''$ (Hayashino et al. 2004; Matsuda et al. 2011; Yamada et al. 2012). We measure an integrated flux density of $S_{870} = 192 \pm 57$ μJy for the source. We stack the nine non-detections in the *uv*-plane to measure the average flux density of these sources. The stacked visi-

Table 1
Target properties.

| ID | RA<br>h m s | Dec<br>° ′ ″ | $z_{\rm spec}$ | $M_\star$<br>$/10^9\,M_\odot$ | $L_{\rm FIR}$<br>$/10^{10}\,L_\odot$ | $L_{\rm UV}$<br>$/10^{10}\,L_\odot$ | IRX | $\beta$ |
|---|---|---|---|---|---|---|---|---|
| SSA22a-C40 | 22 17 19.41 | +00 17 12.7 | 2.927 | $1.0 \pm 0.7$ | $< 2.9$ | $1.8 \pm 0.9$ | $< 1.65$ | $-1.93 \pm 0.21$ |
| SSA22a-C35 | 22 17 20.22 | +00 16 51.9 | 3.103 | $2.7 \pm 1.2$ | $< 3.3$ | $3.7 \pm 1.0$ | $< 0.89$ | $-1.74 \pm 0.12$ |
| SSA22a-C10 | 22 17 20.40 | +00 13 38.5 | 2.812 | $0.9 \pm 0.6$ | $< 3.0$ | $1.2 \pm 0.9$ | $< 2.54$ | $-1.80 \pm 0.26$ |
| SSA22a-C39 | 22 17 20.99 | +00 17 08.9 | 3.076 | $1.3 \pm 0.9$ | $< 2.9$ | $1.7 \pm 1.0$ | $< 1.64$ | $-1.81 \pm 0.23$ |
| SSA22a-C31 | 22 17 22.89 | +00 16 08.9 | 3.023 | $0.6 \pm 0.5$ | $< 1.8$ | $3.5 \pm 1.0$ | $< 0.52$ | $-2.10 \pm 0.11$ |
| SSA22a-C32 | 22 17 25.63 | +00 16 12.4 | 3.301 | $1.9 \pm 1.0$ | $< 4.1$ | $6.5 \pm 1.1$ | $< 0.63$ | $-2.05 \pm 0.08$ |
| SSA22a-C16 | 22 17 31.95 | +00 13 16.1 | 3.065 | $31.6 \pm 3.6$ | $4.8 \pm 1.3$ | $8.4 \pm 0.2$ | $0.56 \pm 0.15$ | $-1.25 \pm 0.03$ |
| SSA22a-C26 | 22 17 39.53 | +00 15 15.1 | 3.178 | $1.6 \pm 1.1$ | $< 3.3$ | $1.3 \pm 1.1$ | $< 2.48$ | $-1.94 \pm 0.28$ |
| SSA22a-C27 | 22 17 43.06 | +00 15 22.1 | 3.084 | $6.5 \pm 3.2$ | $< 3.1$ | $2.6 \pm 1.0$ | $< 1.19$ | $-1.03 \pm 0.32$ |
| SSA22a-C36 | 22 17 46.07 | +00 16 43.3 | 3.066 | $5.7 \pm 2.5$ | $< 3.1$ | $3.5 \pm 1.0$ | $< 0.88$ | $-1.56 \pm 0.22$ |
| Stack | ... | ... | 3.063 | $1.6 \pm 1.1$ | $1.7 \pm 0.5$ | $2.6 \pm 0.2$ | $0.65 \pm 0.21$ | $-1.96 \pm 0.12$ |

**Note.** — The stack consists of the 9 non-detections. Upper limits are quoted at $3\sigma$. Coordinates are Epoch J2000.

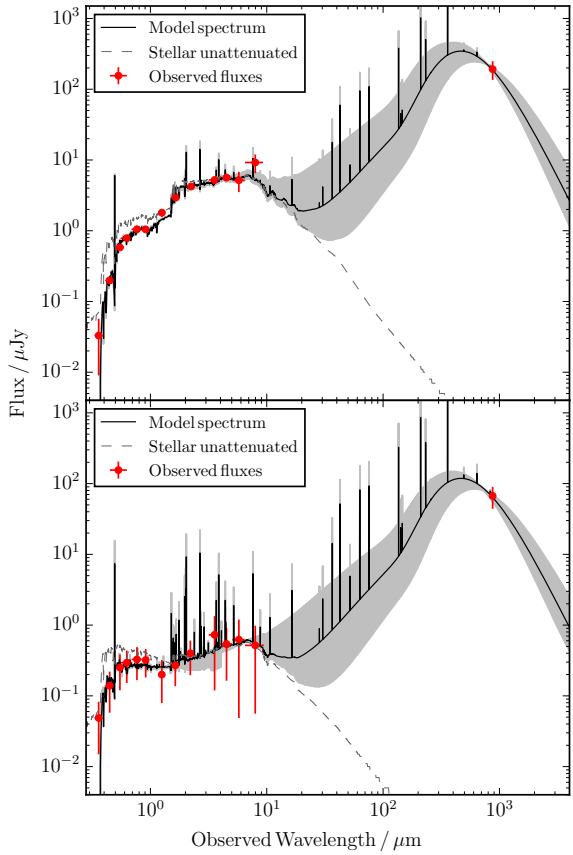

**Figure 3. Top:** The best-fit SED from cigale for SSA22a-C16 at $z = 3.065$, with reduced $\chi^2 = 0.49$ (black solid line), resulting in in integrated IR luminosity $L_{8-1000} = (8.4 \pm 2.3) \times 10^{10}\,{\rm L}_\odot$ (with an additional systematic uncertainty of 0.14 dex, see text for details). **Bottom:** The best-fit SED for the stacked non-detections, giving an average $L_{8-1000} = (2.9 \pm 0.9) \times 10^{10}\,{\rm L}_\odot$. The errors on the photometry represent the standard deviation of the photometry values for the nine galaxies in the stack.

bilities are imaged using same clean procedure as above, but applying a $0.8''$ taper. We detect a $3\sigma$ signal at the phase centre with flux density $S_{870} = 67 \pm 23\,\mu{\rm Jy}$.

As can be seen in Figure 1, SSA22a-C16 has been resolved by ALMA. To confirm this, we plot the average amplitude as a function of baseline separation in Figure 2. An unresolved source has a constant amplitude for all baselines, but the data is better fit by a Gaussian profile (FWHM of $0.94''$). The $\chi^2$ difference between the Gaussian model and a flat profile corresponds to $4.7\sigma$.

## 3. ANALYSIS

### 3.1. *Spectral energy distribution fitting*

We make use of extensive multi-wavelength imaging of SSA22, including CFHT, Subaru (Hayashino et al. 2004; Matsuda et al. 2011; Yamada et al. 2012; Kubo et al. 2013) and *Spitzer*-IRAC (Webb et al. 2009) imaging, to obtain UV through mid-infrared photometry of the targets. We then fit the spectral energy distributions (SEDs) using CIGALE[3] (Noll et al. 2009; Serra et al. 2011). We use stellar population templates from Bruzual & Charlot (2003) with the double-burst star formation history and a Chabrier (2003) initial mass function. Extinction is implemented using Calzetti et al. (2000) and thermal dust emission uses the model of Casey et al. (2012). Since only one photometry point was available in the far-infrared, the mid-infrared power-law slope, $\alpha$, dust emissivity index, $\gamma$, and dust temperature, $T_{\rm d}$, were fixed to 2.0, 1.6 and 37 K respectively. Our choice of $T_{\rm d} = 37\,{\rm K}$ is based on a stacked *Herschel*+SCUBA-2 SED of 1000s of LBGs at $z \sim 3$ from Coppin et al. (2015). In order to estimate the systematic uncertainty on $L_{8-1000}$, we (conservatively) allowed the dust emission parameters to vary between $\alpha = 1.5$–2.5, $\gamma = 1.2$–2.0 and $T_{\rm d} = 27$–47 K, to include more extreme sources (e.g. Saintonge et al. 2013). This resulted in the additional systematic uncertainty of 0.14 dex in the integrated infrared luminosity. The derived physical properties are summarized in Table 1. The best fit SED for SSA22a-C16 is shown in the top panel of Figure 3. For the ALMA non-detections, we averaged the UV-mid-IR photometry and fit the SED in the same way using the stacked ALMA flux. The corresponding best fit for the 'average' LBG is shown in the bottom panel of Figure 3. It is interesting to note that the average mass of the ALMA non-detected LBGs, $M_\star = (1.61 \pm 1.08) \times 10^9\,{\rm M}_\odot$, is a factor twenty lower than that of the LBG we directly detected with ALMA – $M_\star = (3.16 \pm 0.36) \times 10^{10}$ M$_\odot$.

---
[3] http://cigale.lam.fr/

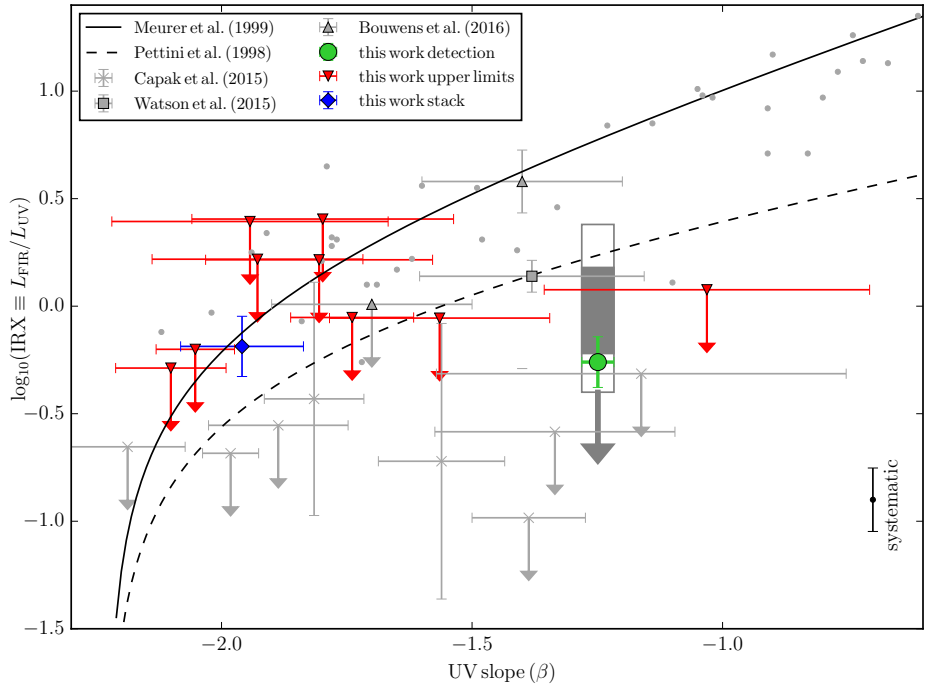

**Figure 4.** IRX as a function of the UV slope $\beta$. Our results are compared with that of Capak et al. (2015) at $z = 5 - 6$, Watson et al. (2015) at $z = 7.5$, Bouwens et al. (2016) at $z \simeq 2 - 3$ and the original Meurer et al. (1999) local sample (with $3\sigma$ upper limits). The lines show the IRX-$\beta$ model of Meurer et al. (1999) for a Calzetti dust law (solid) and SMC-like dust law (Pettini et al. 1998, dashed). The grey filled rectangle depicts the range in IRX measured across our detection (Panel 3 of Figure 1), the grey open rectangle shows the $1\sigma$ systematic uncertainty coming from the relative *HST*-ALMA astrometry uncertainty of $0.2''$ and we also indicate the systematic uncertainty coming from a range of thermal dust emission models. High-$z$ results are clearly highly scattered, and often consistent with very low IRX values. Our resolved map shows that IRX can vary by at least a factor of 3 in a single source, hinting that much of the scatter in the IRX–$\beta$ can be attributed simply to complex relative dust/stellar distributions in star-forming galaxies. A picture is emerging that the starburst nuclei and compact star-forming galaxies used to derive the MHC99 relation sample are poor analogues of the galaxy-integrated values of IRX at high-$z$. Finally, note that, to be consistent with MHC99, we have corrected all points to $L_{\rm FIR}/L_{\rm UV}$, by converting $L_{8-1000}$ or $L_{3-1100}$ (depending on the sample used) to $L_{\rm FIR}$, which typically corresponds to a factor of 0.4–0.5.

### 3.2. *Infrared Excess*

We estimate the SFR for SSA22a-C16, following Madau & Dickinson (2014), with SFR$_{\rm UV} = \mathcal{K}_{\rm UV} \times L_{1500}$ and SFR$_{8-1000} = \mathcal{K}_{8-1000} \times L_{8-1000}$, where $\mathcal{K}_{\rm UV} = 1.15 \times 10^{-28}$ M$_\odot$ yr$^{-1}$ (erg s$^{-1}$ Hz$^{-1}$)$^{-1}$ and $\mathcal{K}_{8-1000} = 4.5 \times 10^{-44}$ M$_\odot$ yr$^{-1}$ (erg s$^{-1}$)$^{-1}$, with $\nu L_{1500} = (8.8 \pm 0.3) \times 10^{10}$ L$_\odot$ and $L_{8-1000} = (8.4 \pm 2.3) \times 10^{10}$ L$_\odot$[4]. We find SFR$_{\rm UV} = 19 \pm 1$ M$_\odot$ yr$^{-1}$ and SFR$_{8-1000} = 15 \pm 4$ M$_\odot$ yr$^{-1}$, giving a total SFR $= 34 \pm 4$ M$_\odot$ yr$^{-1}$. Recall that IRX $\equiv L_{\rm FIR}/L_{\rm UV}$, with $L_{\rm FIR} = (4.8 \pm 1.3) \times 10^{10}$ L$_\odot$ (= $0.57 \times L_{8-1000}$). The corresponding infrared excess is thus (IRX) $= 0.56 \pm 0.15$, and the UV slope (evaluated by fitting the continuum slope of the best-fit SED over rest-frame 1250-2500Å) is $\beta = -1.25 \pm 0.03$. Thus, we can place this galaxy in context with other systems at low and high-$z$ by placing it on the IRX-$\beta$ plot (Figure 4). Averaged over the galaxy it can be seen that our source falls significantly below the MHC99 IRX relation. The stacked detection (and individual upper limits) is more consistent with MHC99, although they are generally bluer and there is still clearly a high degree of scatter in IRX for a fixed $\beta$. One route to understanding the origin of the scatter in IRX-$\beta$ is to use our resolved detection to explore the variation of IRX *within* a single source.

Since we resolve both the rest-frame FIR and UV continuum emission, we can construct a coarse map of IRX. To do this, we grid the ALMA and *HST* images to the same scale and convolve the *HST* image with the ALMA synthesized beam to match it to the lower ALMA angular resolution (we have confirmed that the *HST*/ACS PSF is negligible compared to the ALMA beam). Individual pixel fluxes are converted to IR and UV luminosities using the best-fit (galaxy integrated) SED as a scale. The ratio of the luminosity maps defines the IRX map, and we show this in Figure 1. In regions of the ALMA map without significant submillimeter emission but significant optical emission, the IRX derived is an upper limit. The submillimeter emission sits at the 'saddle' between two clumps of bright UV emission, with the overall morphology resembling a coalescing merger or chain galaxy. Whatever its nature, it is clear that IRX varies strongly across the source (visualised by grey rectangle in Figure 4), with at least a factor of 3 varia-tion from the highly obscured peak of submillimeter emission to the bright knots of UV emission. An important caveat is that we do not have a resolved map of $\beta$ (the F814W band image is our only high-resolution optical image, with the $\beta$ evaluated from the seeing-limited photometry). Resolving $\beta$ would be informative because it would allow us to determine which optical 'component' is dominating the galaxy-averaged UV slope.

---

[4] The Madau & Dickinson (2014) calibration of SFR$_{\rm UV}$ is for 1500Å. However, note that the MHC99 definition of the IRX uses 1600Å.

Finally, to quantify the impact the *HST*-ALMA relative astrometric uncertainty of 0.2″ has on the pixel-to-pixel variations of the IRX, we performed a simple Monte Carlo simulation. We varied the relative positions of the *HST* and ALMA maps 1000 times by shifting the phase center of the ALMA map by a random offset sampled from a gaussian distribution with $\sigma = 0.2''$. For each of the 1000 realizations, the IRX map was constructed in the same way as the 'true' observations and we take the standard deviation of the IRX range measured at the same position as the 'real' map to be the $1\sigma$ systematic error; we depict this in Figure 4 by a grey open rectangle.

## 4. DISCUSSION & SUMMARY

As noted by MHC99, a major caveat in the application of the 'standard' IRX-$\beta$ relation to high-$z$ star-forming galaxies is the assumption that they are analogous to local starbursts (the original sample for which IRX–$\beta$ was derived consists of starburst nuclei, starburst rings, blue compact dwarfs and blue compact galaxies). These systems are characterized by compact star-forming regions in which the starlight is well described by a single stellar population. The dust responsible for obscuring this light is reasonably co-located with the young stars, and so the UV slope is directly associated with the re-radiated thermal dust emission, even assuming a simplistic screen approximation for the dust geometry. In high-$z$ star-forming galaxies the situation is likely to be different: star formation is likely to be more clumpy and widely distributed, driven by interactions and mergers as well as the potentially unstable nature of turbulent gas-rich disks (Ivison et al. 2011; Casey et al. 2014; Simpson et al. 2015). This can result in a complex geometry for the dust with respect to the stars, and so when considering galaxy-integrated properties it is important to note that the shape of the stellar SED is determined by a mix of stellar populations with potentially strongly differentiated extinction.

This appears to be the situation here; the morphology of the rest-frame UV light and the infrared emission are clearly different (in fact, the LBG could be undergoing a merger; see Hine et al. 2016). Although the galaxy has approximately equal amounts of obscured and unobscured star formation, our resolved map of IRX shows that there are regions that are quite obscured (IRX$\simeq$1.5) and regions that are relatively unobscured (IRX$\lesssim$0.5). This scatter in a single system predicts a large scatter in the *unresolved* IRX–$\beta$ for typical star-forming galaxies at high-$z$ for two reasons: (1) if dust and stars are not well mixed, as in this example, then random orientation will play an important role in the observed galaxy-averaged IRX and $\beta$, typically biasing towards higher IRX and redder $\beta$; and, related, (2) short dynamical times could potentially affect the observed IRX–$\beta$ on similar timescales, again driving scatter. Therefore if a single IRX-$\beta$ correction is applied (regardless if it is consistent with MHC99), this large scatter must be considered as an important uncertainty in total SFR estimates if galaxy-integrated values are considered. With this in mind, the question therefore is what the appropriate IRX-$\beta$ correction is for high-$z$ star-forming galaxies? Our galaxy-averaged measurements show that the LBG is not consistent with the MHC99 relation, even when adapted for an SMC-like reddening law.

Bearing in mind the factors described above, evidence is mounting that high-$z$ star-forming galaxies have systematically lower IRX for a fixed $\beta$ than observed for local starburst nuclei (see Figure 4). One explanation put forward, beyond simple geometry arguments, is the nature of interstellar dust at early epochs. At a fundamental level, dust must build up in the ISM over time, such that galaxies at $z \approx 5$ have had less then 1 Gyr to enrich their ISM. Interestingly, Watson et al. (2015) present an ALMA detection of a $z = 7.5$ gravitationally lensed LBG with implied IRX=$1.4 \pm 0.3$, showing that substantial dust reservoirs must be accumulating rather quickly in the first galaxies. It seems unlikely therefore that systematic offsets from the local relation are driven purely by metal abundance. However, without a firmer understanding of both the extinction law *and* the typical joint dust and UV morphology of a larger sample typical high-$z$ star-forming galaxies it will be difficult to disentangle this.

Regardless of the details governing IRX on galactic scales, with direct measurements of both UV and infrared components in high-$z$ galaxies there is growing evidence that the standard MHC99 IRX-$\beta$ relation will typically overestimate the total SFRs of galaxies for a given $\beta$ and this should be an important consideration for any assessment of volume-averaged SFRs based on UV luminosity functions alone. This issue will only be truly resolved with a more extensive survey of the joint IR and UV emission of a large unbiased sample of UV-selected high-$z$ galaxies; a goal we have demonstrated ALMA can achieve.

We thank the referee for a constructive report. We also thank George Bendo from the ALMA UK ARC node for assistance in reducing the ALMA data. K.E.K.C., M.P.K. & N.K.H. acknowledge support from the UK's Science and Technology Facilities Council (grant number ST/M001008/1 & ST/K502029/1). J.E.G. is supported by the Royal Society. K.K. acknowledges support from the Swedish Research Council and the Knut and Alice Wallenberg Foundation. This paper makes use of the following ALMA data: ADS/JAO.ALMA#2013.1.00362.S ALMA is a partnership of ESO (representing its member states), NSF (USA) and NINS (Japan), together with NRC (Canada), NSC and ASIAA (Taiwan), and KASI (Republic of Korea), in cooperation with the Republic of Chile. The Joint ALMA Observatory is operated by ESO, AUI/NRAO and NAOJ.

## REFERENCES

Bouwens R., et al., 2016, ArXiv e-prints (astro-ph/1606.05280)
Bouwens R.J., et al., 2015, ArXiv e-prints (astro-ph/1506.01035)
Bruzual G., Charlot S., 2003, MNRAS, 344, 1000
Calzetti D., Armus L., Bohlin R.C., Kinney A.L., Koornneef J., Storchi-Bergmann T., 2000, ApJ, 533, 682
Capak P.L., et al., 2015, Nature, 522, 455
Casey C.M., et al., 2012, ApJ, 761, 140
Casey C.M., Narayanan D., Cooray A., 2014, Phys. Rep., 541, 45
Chabrier G., 2003, ApJ, 586, L133
Coppin K.E.K., et al., 2015, MNRAS, 446, 1293
Douglas L.S., Bremer M.N., Stanway E.R., Lehnert M.D., Clowe D., 2009, MNRAS, 400, 561
Ellis R.S., et al., 2013, ApJ, 763, L7
Hayashino T., et al., 2004, AJ, 128, 2073
Helou G., Khan I.R., Malek L., Boehmer L., 1988, ApJS, 68, 151
Hine N.K., Geach J.E., Alexander D.M., Lehmer B.D., Chapman S.C., Matsuda Y., 2016, MNRAS, 455, 2363
Ivison R.J., Papadopoulos P.P., Smail I., Greve T.R., Thomson A.P., Xilouris E.M., Chapman S.C., 2011, MNRAS, 412, 1913

Kubo M., et al., 2013, ApJ, 778, 170
Lawrence A., et al., 2007, MNRAS, 379, 1599
Madau P., Dickinson M., 2014, ARA&A, 52, 415
Matsuda Y., et al., 2011, MNRAS, 410, L13
McLeod D.J., McLure R.J., Dunlop J.S., 2016, ArXiv e-prints
Meurer G.R., Heckman T.M., Calzetti D., 1999, ApJ, 521, 64
Noll S., Burgarella D., Giovannoli E., Buat V., Marcillac D., Muñoz-Mateos J.C., 2009, A&A, 507, 1793
Oesch P.A., et al., 2014, ApJ, 786, 108
Oesch P.A., Bouwens R.J., Illingworth G.D., Franx M., Ammons S.M., van Dokkum P.G., Trenti M., Labbé I., 2015, ApJ, 808, 104
Oke J.B., et al., 1995, PASP, 107, 375
Pettini M., Kellogg M., Steidel C.C., Dickinson M., Adelberger K.L., Giavalisco M., 1998, ApJ, 508, 539
Saintonge A., et al., 2013, ApJ, 778, 2
Serra P., Amblard A., Temi P., Burgarella D., Giovannoli E., Buat V., Noll S., Im S., 2011, ApJ, 740, 22
Simpson J.M., et al., 2015, ApJ, 799, 81
Steidel C.C., Adelberger K.L., Shapley A.E., Pettini M., Dickinson M., Giavalisco M., 2003, ApJ, 592, 728
Swinbank A.M., et al., 2015, ApJ, 806, L17
Tacconi L.J., et al., 2010, Nature, 463, 781
Watson D., Christensen L., Knudsen K.K., Richard J., Gallazzi A., Michałowski M.J., 2015, Nature, 519, 327
Webb T.M.A., Yamada T., Huang J.S., Ashby M.L.N., Matsuda Y., Egami E., Gonzalez M., Hayashimo T., 2009, ApJ, 692, 1561
Yamada T., Nakamura Y., Matsuda Y., Hayashino T., Yamauchi R., Morimoto N., Kousai K., Umemura M., 2012, AJ, 143, 79